\documentclass{gretsi}

\usepackage[latin1]{inputenc}          % UNIX, codage ISO 8859-1
\usepackage[english,french]{babel}     % "babel.sty" + "french.sty"
\usepackage{times}		         % ajout times le 30 mai 2003 
\usepackage[T1]{fontenc}

\usepackage{amsmath,amssymb}
\usepackage{graphicx}

\title{Wavelet graphs for the direct detection of gravitational waves}

\author{\coord{Eric}{Chassande-Mottin}{1},
        \coord{Eric}{Lebigot}{1,2},
    \coord{Hugo}{Magaldi}{1},
    \coord{Eve}{Chase}{1},
    \coord{Archana}{Pai}{3},
    \coord{Gayathri}{V}{3},
    \coord{Gabriele}{Vedovato}{4}
  }

\address{\affil{1}{APC, Univ Paris Diderot, CNRS/IN2P3, CEA/Irfu, Obs. de Paris, Sorbonne Paris Cité, France}
         \affil{2}{Tsinghua University, Beijing, China}
         \affil{3}{IISERTVM, Computer Science Building, CET Campus, Trivandrum Kerala, India}
         \affil{4}{INFN, Sezione di Padova, Padova, Italia}
}

\email{ecm@apc.univ-paris7.fr}

\frenchabstract{Une seconde génération de détecteurs d'ondes gravitationnelles
  entrera prochainement en fonction avec l'objectif de mesurer pour la première
  fois le faible signal gravitationnel provenant de la coalescence de binaires
  de trous noirs et/ou d'étoiles à neutrons.  Dans cette communication, nous
  proposons une méthode de recherche temps-fréquence alternative au filtre
  adapté habituellement utilisé pour détecter ce signal. Cette méthode repose
  sur l'utilisation d'un graphe qui encode, en créant des liens entre les
  coefficients de la décomposition temps-fréquence multi-échelle des données,
  l'évolution temporelle du signal ainsi que sa variabilité. Nous donnons une
  preuve de concept de l'approche proposée.}

\englishabstract{A second generation of gravitational wave detectors will soon
  come online with the objective of measuring for the first time the tiny
  gravitational signal from the coalescence of black hole and/or neutron star
  binaries. In this communication, we propose a new time-frequency search method
  alternative to matched filtering techniques that are usually employed to detect
  this signal. This method relies on a graph that encodes the time evolution of
  the signal and its variability by establishing links between coefficients in the
  multi-scale time-frequency decomposition of the data. We provide a proof of
  concept for this approach.}

\begin{document}

\maketitle

\section{Context and motivation}

Einstein's theory of General Relativity introduces the concept of a deformable
and evolving space-time. The dynamics of space-time is prescribed by Einstein's
equations. In the linearised gravity framework (space-time metric is a small
perturbation to the Minkowsian flat space-time metric), the Einstein's equations
can be transformed into the wave equation. The metric perturbation evolve and
propagate like radiation with amplitude scaled as 1/r and travels with speed of
light. These are referred to as \textit{gravitational waves} (GW) as they
propagate as disturbances of space-time itself
\cite{thorne87:_gravit_radiat}. GWs have never been directly detected so far,
i.e., through the measurement of their effect on a man-made instrument.

The direct search for GWs has made significant progress with the advent of
dedicated instruments based on high-precision laser interferometry. A worldwide
network of kilometric-scale interferometric GW detectors including the US-based
LIGO\cite{aasi15:_advan_ligo}, the French-Italian project
Virgo\cite{acernese15:_advan_virgo} completed a first series of science data
collection over the past decade and will soon resume to take data in an advanced
(ten times more sensitive) configuration. The first discovery of GWs is expected
within the decade; this will open an entirely new view of the universe.

Those instruments are designed to sense the tiny space-time strain distorsion
inside the detector enclosure caused by GW from distant astrophysical sources.
Coalescing binaries of neutron star and/or black holes (in short, CBCs for
compact binary coalescences) are one of the most promising sources of GW. The
last minutes before the binary merges coincide with the emission of an intense
burst of GWs. An accurate modeling of the dynamics of the binary shows that the
GW waveform is a quasi-periodic signal, or chirp. The chirp frequency sweeps
towards high values according to a power law. The baseline approach is to use
this morphological information to search the data with matched filtering
techniques (see e.g. \cite{chassande-mottin12:_data}).

A large amount of computational resources is required to complete the search.
This results from the combined effect of the large volume of data and physical
signal parameter space to be searched and the impossibility to accurately model
the instrumental noise. This implies that the analysis background has to be
estimated empirically by repeating the analysis many (typ. $10^5$) times on
surrogate data obtained by shifting data streams with non-physical time delays.

In this article we propose a new wavelet-based method, alternative to matched
filtering, to search for CBC signals. It relies on the multi-scale
representation of CBC chirps using wavelet graphs (introduced in
Sec.~\ref{sec:waveletgraph}). The method is integrated in the data analysis
pipeline Coherent WaveBurst that we present in the next section.

\subsection{Multi-scale coherent transient searches}
\label{sec:coherentWB}

A whole range of data analysis pipelines has been developed for searching for GW
transients (including CBC signals) through a combined analysis of multiple
detector data. Among these, an important pipeline is Coherent WaveBurst (cWB)
\cite{klimenko05:_const}.

In a nutshell cWB extracts clusters of significant coefficients from
time-frequency decompositions that results from the \textit{coherent}
combination of the data using sensor array techniques analog to beam-forming.
The resulting clusters form candidate GW ``events'' if their ``coherent'' to
``incoherent'' signal energy ratio exceeds a threshold.  We now detail how the
time-frequency representations is computed and how the clusters are formed.

\subsubsection{Multi-scale Wilson transform}
\label{sec:multiscale_Wilson}

In cWB's scheme, the data (time series) are mapped into a set of time-frequency
representations by projecting onto Wilson bases
\cite{daubechies92:_ten_lectur_wavel} which are a variation of the well-known
Gabor bases. A Wilson basis is composed of linear phase cosine modulated
wavelets distributed on a regular time-frequency lattice. According to
\cite{necula12:_trans_wilson_daubec} a pair of in-phase and quadrature
orthonormal Wilson bases is constructed. Time-frequen\-cy maps are computed by
summing the powers in the in-phase and quadrature transforms. A collection of
maps is computed using the Meyer scaling function
\cite{necula12:_trans_wilson_daubec, daubechies92:_ten_lectur_wavel} with
different \textit{scales} (i.e., wavelet duration or time scale). At scale $a$,
the time-frequency lattice includes $a+1$ frequency subbands and $N/a$ time bins
where $N$ is the number of data samples (typ., $N=2^{20}$ i.e, $1024$ s duration
sampled at $f_s = 1024$ Hz).

The collection of time-frequency at all scales results in a three-dimensional
redundant representation that spans the $t, f$ and $a$ variables. In the cWB
scheme, the selected $N_a$ scales are distributed dyadically, typ. $\log_2 a \in
\{ 3, \ldots, 8\}$.  The total number of $(t,f,a)$ pixels is $\sim N_a N \approx
6 \times 10^6$.

\subsubsection{Transient extraction from Wilson transform}

Significant pixels in the time-frequency-scale representation are selected by
thresholding. In each time-frequency maps, the selected pixels are clustered
using a nearest neighbourgh algorithm. The time-frequency clusters obtained at
all scales are then combined using an algorithm that selects the principal
components. This procedure does not make any prior assumption on the cluster
geometry which can have an arbitrary shape in principle.

Because of their specific phase evolution, chirp signals have structured
time-frequency representations with energy mostly concentrated on the
instantaneous frequency curve (see e.g., \cite{delprat92:_asymp_gabor}) thus
leading to clusters of significant pixels with a specific shape. \textit{We
  propose here a new clustering algorithm that targets that shape}.

\section{Wavelet graphs}
\label{sec:waveletgraph}

In this section, we determine the time-frequency-scale curve referred to as
\textit{chirp path} that collects the large wavelet coefficients associated to a
given chirp signal. In presence of stationary noise, the coefficients in the
path are the one which maximizes the signal-to-noise ratio locally.

We construct a graph that combines the paths from a family of chirp signals
that covers a region of the parameter space. This wavelet graph is a central
piece of the clustering algorithm we propose.

\subsection{Chirp expansion in Wilson bases}
\label{sec:chirppath}

For a given analysis frequency, we now determine which wavelet in the Wilson
basis has the maximum coupling with the considered chirp.

Instead of using discrete Wilson transforms with Meyer wavelets as cWB, we
work in the continuous time, frequency and scale limit and use sine Gaussian
wavelets to allow analytical calculations. The wavelet at time $t_0$, frequency
$f_0$ and scale $a_0$ reads $\tilde{w}_0(f)=\tilde{g}(f-f_0;a_0) \exp -2\pi i f
t_0$ where $g(\cdot)$ is the wavelet envelope assumed to be
\begin{equation}
\tilde{g}(f;\sigma_0)=(2 \pi)^{1/4} \sqrt{\sigma_0} \exp -\pi^2 \sigma_0^2 (f-f_0)^2.
\end{equation}

The scale parameters in the discrete/Meyer and continuous/\-Gaussian cases are
approximately related by $a_0 \approx f_s \sigma_0$ where $f_s$ is the sampling
frequency.

The time-frequency map is defined as
\begin{equation}
\label{eq:def-wavelet-transform}
\rho^2_0 \equiv \rho^2(t_0, f_0, a_0) = \left| \int df \frac{\tilde{w}_0^*(f)\tilde{s}(f)}{\tilde{N}(f)} \right|^2
\end{equation}
where $\tilde{N}(f)$ is the noise power spectrum. We seek the time $t_0$ and
scale $a_0$ (or equivalently $\sigma_0$) which maximize $\rho_0$ for a given $f_0$.

Chirps (including CBC signals) can be expressed as $\tilde{s}(f)=A(f) \exp i
\Psi(f)$ in the Fourier domain. This allows us to rewrite
Eq.~(\ref{eq:def-wavelet-transform}) as an oscillatory integral. We then
evaluate this integral with the stationary phase approximation
\cite{dalcanton14:_effec_gauss} assuming slow variations of the integrand
amplitude with respect to its phase. We obtain:
\begin{equation}
\label{eq:rho1}
\rho_0^2 \approx  \frac{\pi |{\cal A}(f_0)|^2}{\left| \pi^2 \sigma_0^2 - i \beta \right|} 
\exp \left[\Re \frac{\pi^2 (t_0 - \tau(f_0))^2}{\pi^2 \sigma_0^2 - i \beta} \right],
\end{equation}
where ${\cal A}(f) = (2 \pi)^{1/4} \sqrt{\sigma_0} A(f)/\tilde{N}(f)$,
$\beta=\ddot{\Psi}(f_0)/2$ and $\tau(f_0) = -(2\pi)^{-1} \dot{\Psi}(f_0)$ denotes
the inverse chirp rate and group delay, resp.

The maximization of this quantity in $t_0$ and $\sigma_0$ at the given
frequency yields
\begin{equation}
\label{eq:rho3}
\rho^2(\hat{t}_0, f_0, \hat{a}_0)  = \frac{f_s}{\sqrt{\pi}} \frac{|A(f_0)|^2}{\hat{a}_0 \tilde{N}^2(f_0)},
\end{equation}
with the maximum reached at $\hat{t}_0=\tau(f_0)$ and $\hat{\sigma}_0 =
\sqrt{|\beta|}/\pi$ converted into $\hat{a}_0$ using the Gaussian to Meyer scale
conversion stated above.

We conclude that the chirp path is
the following curve in the time-frequency-scale space parametrized by the
frequency $f_0$
\begin{align}
\label{eq:chirppath}
\hat{t}_0 &=-\frac{1}{2\pi}\dot{\Psi}(f_0) &
\hat{a}_0 &=\frac{f_s}{\sqrt{2}\pi} \sqrt{|\ddot{\Psi}(f_0)|}.
\end{align}

The last expression implies that signals with slowly (resp. rapidly) varying
frequency are best approximated by wavelets of large (resp. small) scale as
expected intuitively.

The chirp path essentially depends on the chirp Fourier phase $\Psi(\cdot)$.
For CBC chirp signals, this phase in the \textit{Newtonian} approximation is
\cite{thorne87:_gravit_radiat}
\begin{equation}
\label{eq:newton_phase}
\Psi(f)=\psi_c - 2\pi f t_c - \frac{6 \pi f_L \tau_0}{5} \left(\frac{f_L}{f}\right)^{5/3} ,
\end{equation}
where $\psi_c$ is the final phase at coalescence time $t_c$. The chirp duration
$\tau_0$ from the lower cut-off $f_L$ to maximum frequency reads
\begin{equation}
\label{eq:chirp time}
\tau_0 = \frac{5}{256} \left(\frac{c^3}{G \mathcal{M}} \right)^{5/3} (\pi f_L)^{-8/3}.
\end{equation}
where the \textit{chirp mass} $\mathcal{M}=(m_1 m_2)^{3/5}/(m_1+m_2)^{1/5}$
depends on the binary component masses $m_1$ and $m_2$.

Eqs.~(\ref{eq:chirppath}) thus leads to
\begin{align*}
\label{eq:chirppath_newton}
\hat{t}_0 &=t_c- \tau_0 \left(\frac{f_0}{f_L}\right)^{-8/3} &
\hat{a}_0 &=\frac{4 f_s}{\sqrt{6~\pi}} \left(\frac{\tau_0}{f_0}\right)^{1/2} \left(\frac{f_0}{f_L}\right)^{-4/3}.
\end{align*}

We verify that the signal enters the detector frequency band ($f_0=f_L$) at the
arrival time $\hat{t}_0= t_c - \tau_0$ and leaves it close to coalescence time.

Eqs.~(\ref{eq:chirppath}) provide approximations in the continuous limit. This
curve has to be discretized according to the $(t, f, a)$ lattice adopted by
cWB. Discrete time and frequency coordinates read $\bar{t}_0 = \lfloor
\hat{t}_0/\delta \rfloor \delta$ and $\bar{f}_0 = \lfloor \hat{f}_0 \delta
\rfloor /\delta$ and are obtained using the time sampling step $\delta =
\bar{a}_0/f_s$ at the discrete scale $\log_2 \bar{a}_0 = \lfloor \log_2
\hat{a}_0 \rfloor$ where $\lfloor \cdot \rfloor$ is the round-off operator. This
results in a finite and ordered pixel collection with coordinates
$(\bar{t}_0, \bar{f}_0, \bar{a}_0)$ we refer to as \textit{chirp path}.

\subsection{Cover chirp space with a graph}

In the above model of Eq.~(\ref{eq:newton_phase}), the coalescence time $t_c$
and component masses $m_1$ and $m_2$ are not known \textit{a priori} and have to
be estimated from the data. For that reason, the chirp paths associated to the
time and mass parameter space are computed and combined into the \textit{wavelet
  graph} that collects the selected pixels and their connection with the
previous pixel in the path (if any), referred to as \textit{ancestor}.  The last
pixel of all paths is marked as an \textit{end node}. If a pixel occurs in two
(or more) chirp paths, the graph retains the list of all its ancestors.

\begin{figure}[htb]
   \begin{center}
     \includegraphics[width=.8\columnwidth]{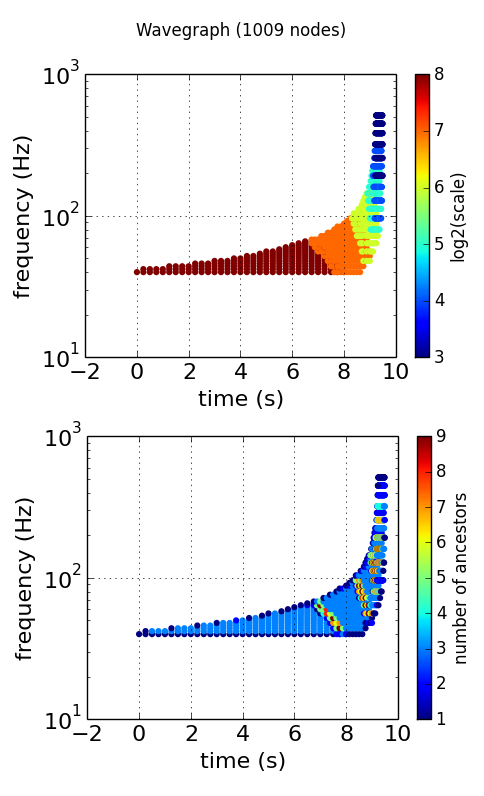}
   \end{center}
  \legende{\label{fig:graph} Typical wavelet graph computed for CBC signal from
    binaries in the mass range $m_1,m_2 \in [2.5, 10] M_{\odot}$ (expressed in
    unit of solar mass). The top panel shows the distribution of
    selected pixel nodes in the $(t,f,a)$ space. The bottom panel shows the
    number of ancestors per node.}
\end{figure}

Fig.~\ref{fig:graph} shows a typical wavelet graph computed for the mass range
$m_1,m_2 \in [2.5, 10] M_{\odot}$ (expressed in unit of solar mass)
and $t_c-t_{\mathrm{ref}} \in [0, \delta_{\mathrm{max}}]$ with
$\delta_{\mathrm{max}} = a_{\mathrm{max}}/f_s \approx 250$ ms with the standard
cWB settings (see Sec.~\ref{sec:multiscale_Wilson}). We fix the analysis
frequency bandwidth from 40 Hz to the Nyquist frequency.

Despite the reasonably large physical space covered (1275 different CBC signals
were used in this computation), the wavelet graph has a moderate size ($\sim$
1000 nodes) and complexity ($\lesssim$ 10 ancestors per node). 

\section{Clustering with wavelet graphs}
\label{sec:clustering}

We now explain how the graph introduced in the previous section can be used to
detect chirps in the data. 

Assuming Gaussian noise, the detection of a known chirp signal can be performed
optimally using matched filtering. The matched filtering statistics can be
re-expressed in the wavelet domain. Assuming that the large coefficients of the
chirp wavelet tranform are essentially contained in the chirp path $p$ obtained
in Sec.~\ref{sec:chirppath} and that the selected wavelets in the chirp path are
nearly orthogonal, this results in:
\begin{equation}
\ell(p) = \sum_{t,f,a \in p} \rho^2(t,f,a),
\end{equation}
where the summation runs from the path start node (no ancestor) to the end node.

\begin{figure*}
  \center
    \includegraphics[width=.85\textwidth]{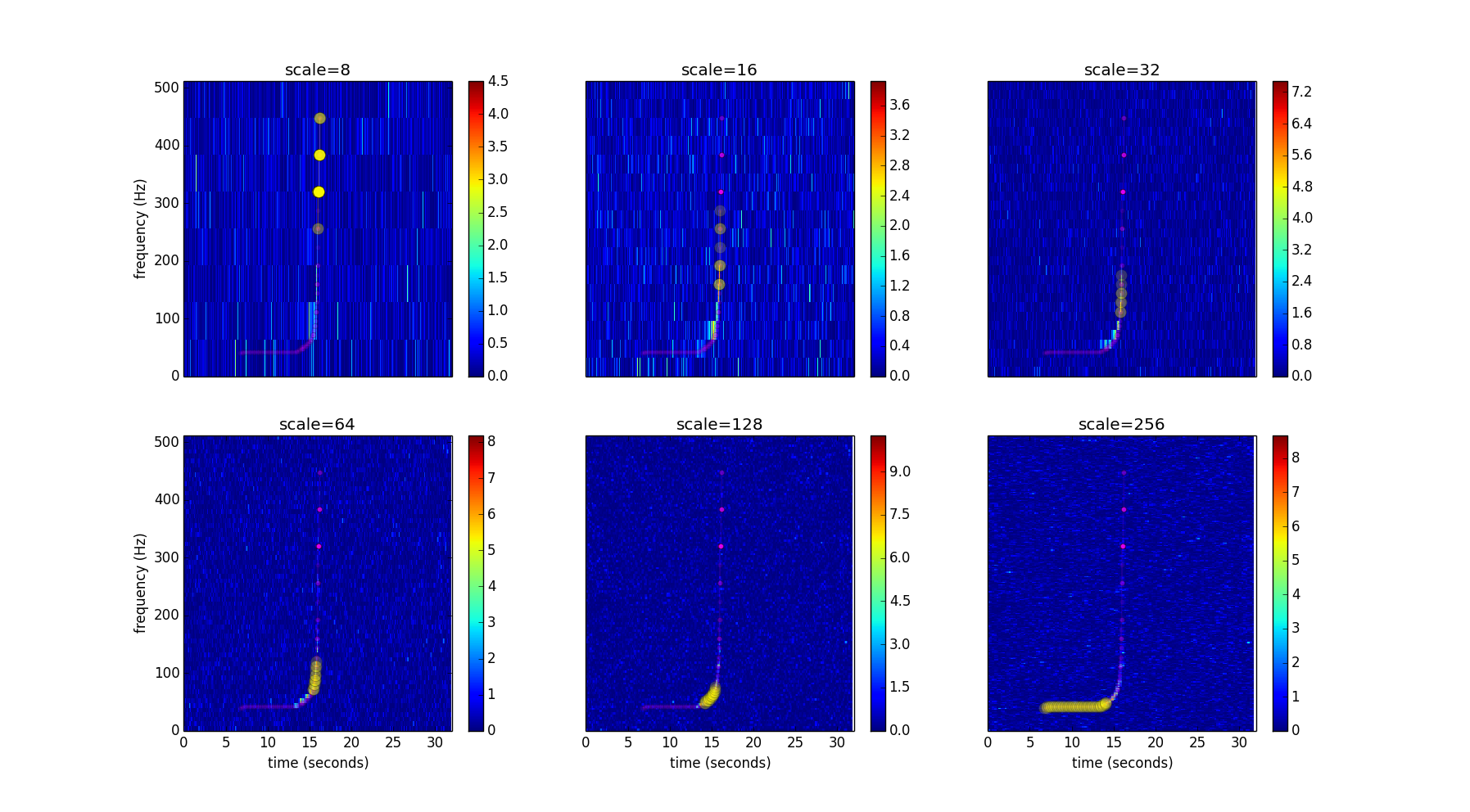}
    \legende{\label{fig:example} Application of the proposed clustering method
      to a noisy CBC signal with $m_1=m_2=5 M_{\odot}$ using the wavelet graph
      in Fig.~\ref{fig:graph}. The extracted cluster is shown in magenta with
      pixels at corresponding scale evidenced with yellow dots.}
\end{figure*}

When the chirp parameters are unknown, this statistics has to be maximized over
the admissible parameter space ${\cal P}$, namely $\max_{p \in {\cal P}}
\ell(p)$.  This maximization amounts to finding the chirp path in the wavelet
graph that captures the largest amount of energy. That can be efficiently
performed using combinatorial optimization techniques such as dynamic
programming with a computing cost scaling linearly with the size of the graph.

We divide the data stream into successive segments and compute their Wilson
transform. We assign to the graph node the values of the corresponding
coefficients in the Wilson transform, apply dynamic programming to extract the
``best'' chirp path and move to the next segment. Chirp paths with $\ell$
exceeding a pre-defined threshold are retained and given to cWB as interesting
clusters for further processing.

\section{Concluding remarks}
\label{sec:conclusion}

As an illustration, Fig.~\ref{fig:example} presents the result of the wavelet
graph clustering method on a CBC chirp signal in Gaussian white noise at large
SNR $\sim 60$. The cluster being continuous across times, frequencies and scales
by design, it collects marginally significative pixels (because of noise
fluctuations) which are lost otherwise in the standard cWB scheme.

Thanks to moderate graph size, the overall computing cost is in the acceptable
range for production. On-going simulations will allow a full evaluation, beyond
the present proof of concept.

\subsection*{Acknowledgements}

{\footnotesize We thank Sergey Klimenko for the access to the coherent WaveBurst
  software and Francesco Salemi for his useful comments on this work. We
  acknowledge the National Science Foundation (Award 1005036 through the
  University of Florida International REU for Gravitational Waves) and CNRS
  (PICS) for their support.}

%\bibliographystyle{plain}
%\bibliography{reference}

\renewcommand{\refname}{References}

% \begin{figure*}
%   \centering
%   \includegraphics[width=\columnwidth]{figure/CBC_example.png}
%  \legende{\label{fig:chirp} Example of a CBC chirp signal with a sketch of its
%    path as obtained in Eq.~(\ref{eq:chirppath}) evidencing few wavelets of
%    long, medium and short durations.}
% \end{figure*}

\end{document}